\documentclass[traditabstract]{aa}
\usepackage{natbib} 
\usepackage{graphicx}
\usepackage{txfonts}

\newcommand{\Teff}  {\mbox{T$_\mathrm{eff}$}\,}
\newcommand{\FeH}   {\mbox{[Fe/H]\,}}
\newcommand{\logg}  {\mbox{$\log$ g}\,}
\newcommand{\tgm}   {\mbox{(T$_\mathrm{eff}$, $\log$g, [Fe/H])} \,}
\newcommand{\pastel}{{\tiny PASTEL}}

\begin{document}

\title{The PASTEL catalogue: 2016 version\thanks{The \pastel\ catalogue is available in electronic form at the Centre de Donn\'ees Stellaires in Strasbourg  (http://vizier.u-strasbg.fr/viz-bin/VizieR?-source=B/pastel)}}
\titlerunning{\pastel\ 2016}

\author{Caroline Soubiran, Jean-Fran{\c c}ois Le Campion, Nathalie Brouillet, \and Laurent Chemin}

\offprints{C. Soubiran, \email{caroline.soubiran@u-bordeaux.fr}}

\institute{Laboratoire d'Astrophysique de Bordeaux, Univ. Bordeaux, CNRS, UMR5804, F-33615, Pessac, France.}
\date{Received 11 March 2016 / Accepted 4 May 2016}

\abstract{ 
 The bibliographical compilation of stellar atmospheric parameters \tgm relying on high-resolution, high signal-to-noise  spectroscopy started in the eighties with the so-called \FeH catalogue, and was continued in 2010 with the \pastel\ catalogue,  which also includes determinations of \Teff alone, based on various methods. Here we present an update of the \pastel\ catalogue. The main journals and the CDS database have been surveyed to find relevant publications presenting new determinations of atmospheric parameters. As of February 2016, \pastel\ includes  64\,082 determinations of either  \Teff or \tgm for 31\,401 stars, corresponding to 1\,142 bibliographical references. Some 11\,197 stars have a determination of the three parameters \tgm with a high-quality spectroscopic metallicity.
}

                
\maketitle
                

\section{Introduction}

 The knowledge of stellar atmospheric parameters \tgm is crucial in many reasearch areas related to the physics of stars and galaxies. 
In particular the atmospheric parameters are needed before the chemical composition of a star can be estimated.  

\pastel\ is a bibliographical catalogue compiling determinations of stellar atmospheric parameters  \citep{sou10}. It provides \tgm determinations obtained from detailed analyses of high-resolution (R$\ge$25\,000), high signal-to-noise spectra (S/N $\ge$ 50), carried out with the help of model atmospheres. \pastel\ is the continuation of the effort made by Giusa Cayrel de Strobel to make high-quality iron abundances available to a wide community in a convenient way. Six editions of the \FeH catalogue have been published \citep{cay80, cay81,cay85, cay92, cay97, cay01}. \pastel\ follows the same philosophy with one modification: it also provides effective temperatures \Teff determined from various methods. We decided to include \Teff because  it is a critical parameter in spectroscopic analyses.  Knowing  \Teff precisely leads to lower uncertainties in abundances \citep[see e.g.][]{jof14,jof15}. It is thus helpful to gather studies providing precise determinations of \Teff, even if not based on high-resolution spectroscopy. In particular, the increase in the accuracy of angular diameter measurements of stars by interferometry has provided a number of direct measurements of \Teff through the fundamental relation, very important because largely model-independent. We found it relevant to include them in \pastel.

\pastel\ is available through VizieR at CDS. It is a popular catalogue, queried about 15\,000 times per month on average. The initial version \citep{sou10} included 30\,151 records. Two updates have been provided, in 2011 (31\,724 records) and 2013 (52\,045 records). We present in this paper the new version of \pastel, as of Febuary 2016, which now has  64\,082 records. Section 2 describes the contents of the revised release of \pastel\ in terms of stellar parameters. Section 3 presents its stellar content.

\section{Description of the catalogue}
\subsection{Star names, coordinates, and magnitudes}
Only stars with known equatorial coordinates in Simbad  are listed in \pastel. Owing to this restriction, several studies could not be included, especially in the regions of stellar clusters where the stars are often ambiguously identified. This restriction, however, ensures that \pastel\ can be cross-matched to other catalogues. 

The Simbad database has recently been cleaned and some stars that were in previous versions of \pastel\ have been removed because of their unknown or too uncertain coordinates.  Fifty-two stars not identified in Simbad can be found in the 2MASS catalogue \citep{cut03} and so the 2MASS coordinates have been adopted.

Like in the previous versions of \pastel, some spectroscopic binaries remain problematic because their parameters could be determined separately from high spectral resolution spectroscopy, while only one entry is available for both components in Simbad. These stars are not included in \pastel. 

For each star recorded in \pastel, Simbad was queried to retrieve its equatorial coordinates (ICRS, epoch J2000) and its B, V,  J, H, K magnitudes. The catalogue can thus be searched by position on the sky and magnitude interval. There are 31\,401 different stars in the current version of \pastel. 

\subsection{Effective temperature \Teff}
 
\begin{figure*}[t!]
\center
\includegraphics[width=0.3\textwidth]{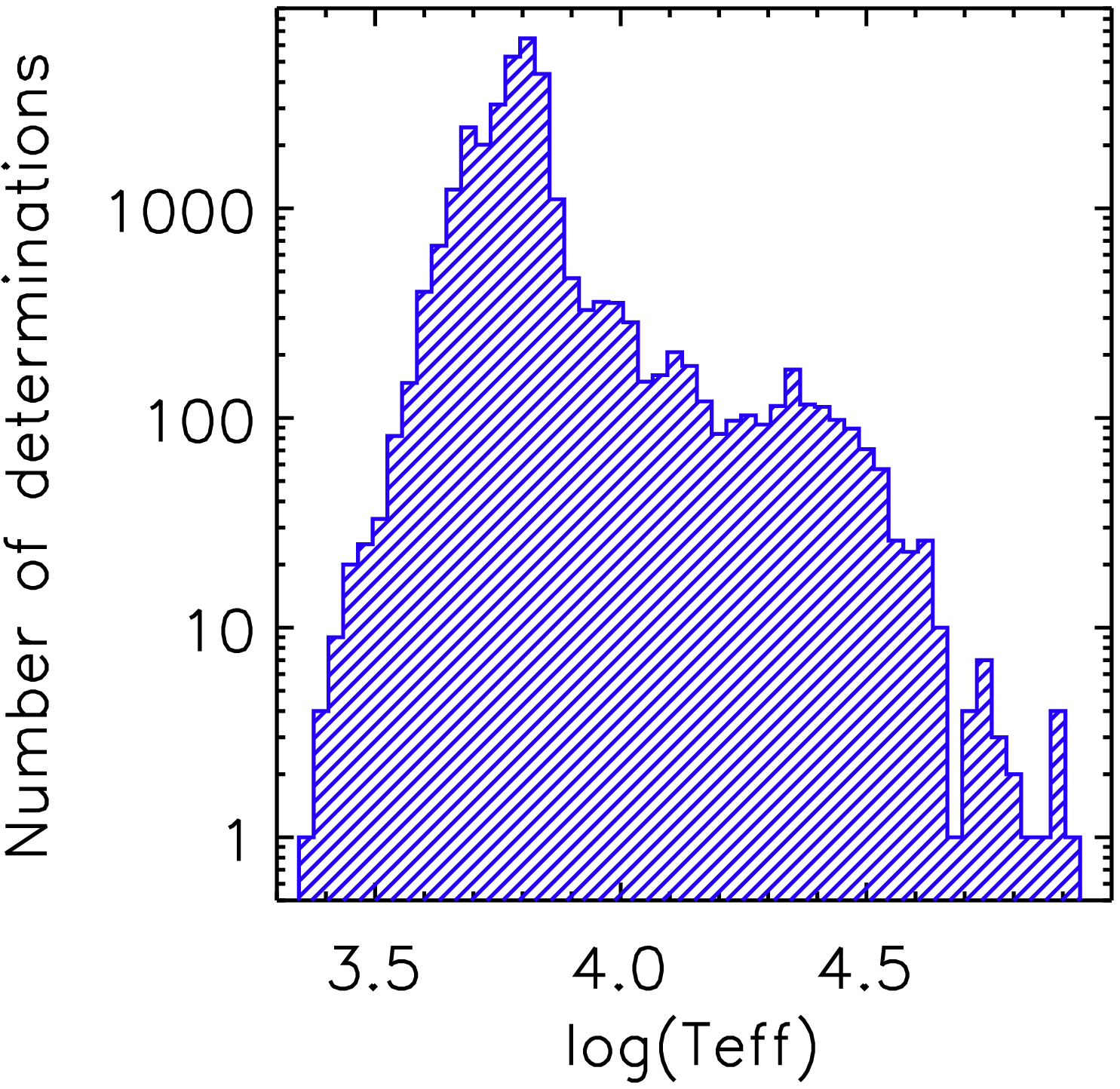}\includegraphics[width=0.3\textwidth]{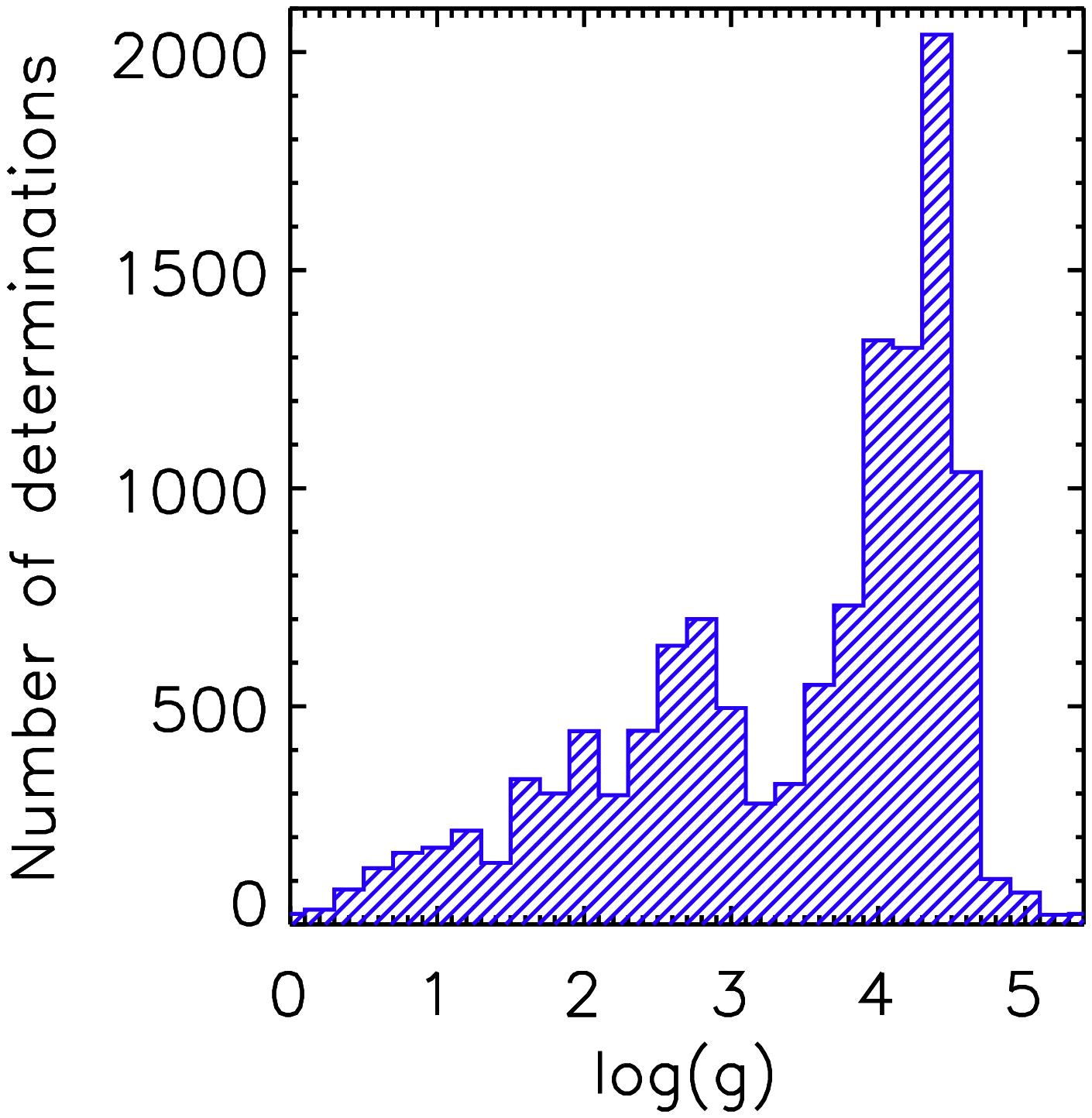}\includegraphics[width=0.3\textwidth]{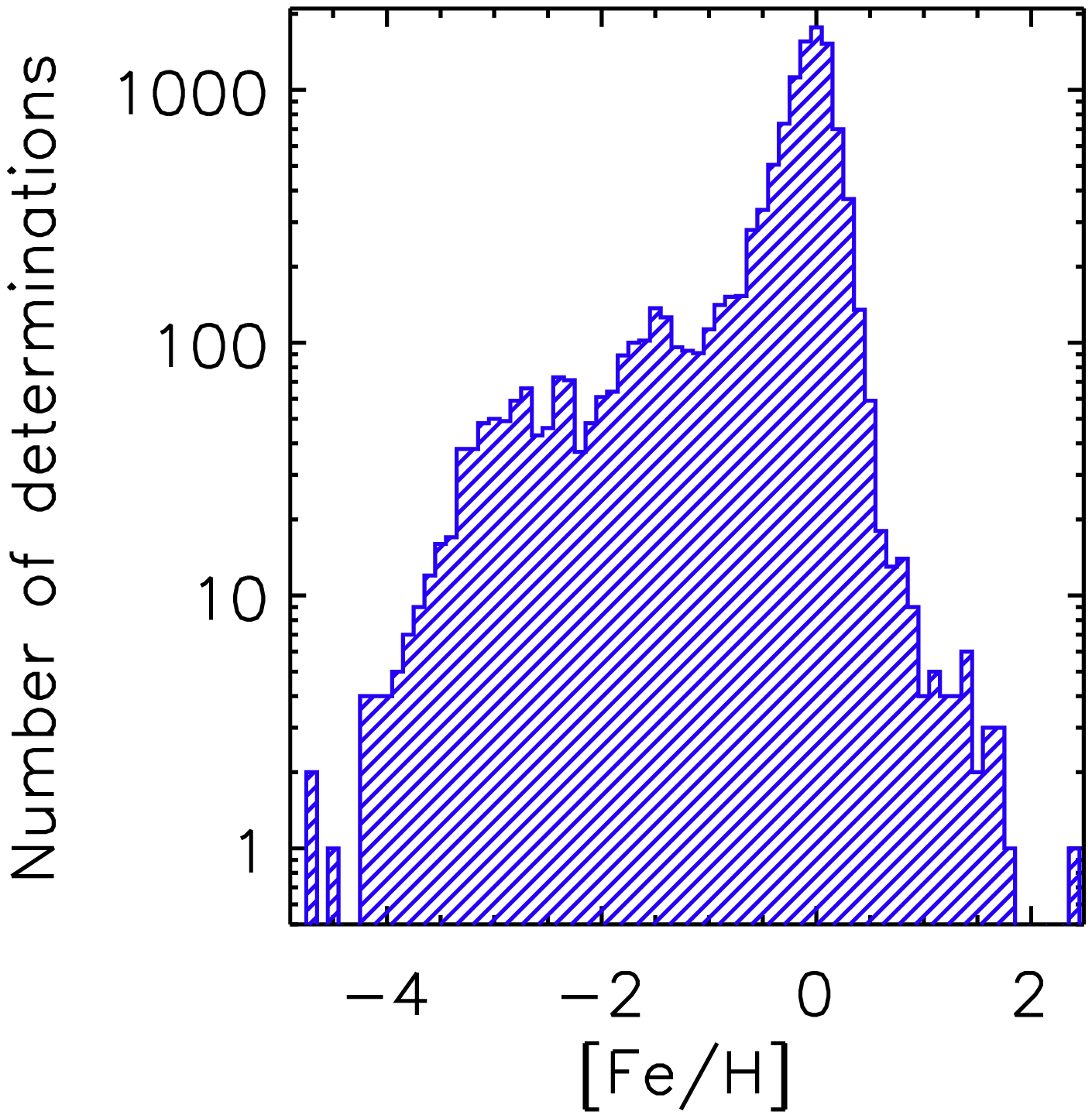}
\caption{Histogram of effective temperatures, surface gravities, and metallicities in \pastel. }
\label{tgm}
\end{figure*}

The 63\,797 \Teff determinations in \pastel\ span a wide range from $\sim$2300 K to $10^5$ K with a median value of 6000 K. The histogram is shown in Fig.~\ref{tgm}.  The vast majority of \Teff determinations correspond to FGK stars. 

If quoted in the articles, uncertainties on \Teff are provided. They vary from less than 5 K for some FGK stars \citep[e.g.][]{kov06,ram14} to $\sim$5\,000 K for some OB  stars \citep[e.g.][]{sok95,lat13}. Translated into relative errors, their median value is 1.1\%. It is worth noting that there is a variety of definitions or procedures to evaluate \Teff uncertainties that are not homogeneous.  When a star has  multiple \Teff values, the standard deviation shows how these different determinations agree with one other, and reflects both internal and external errors, including differences in temperature scales. Nearly half of the stars in \pastel\ have at least two \Teff values from which the standard deviation can be computed. Its median value is 0.76\%, demonstrating that the authors agree quite well in general, even if they have not used  the same observing material and methods. There are outliers, however; some stars have \Teff determinations in significant disagreement (83 and 343 stars have a standard deviation larger than 10\% and 5\%, respectively). To discard discrepant determinations, it is safer to adopt a median instead of an arithmetic mean if an averaged \Teff value is needed. 

\begin{figure}[t]
\center
\includegraphics[width=0.5\textwidth]{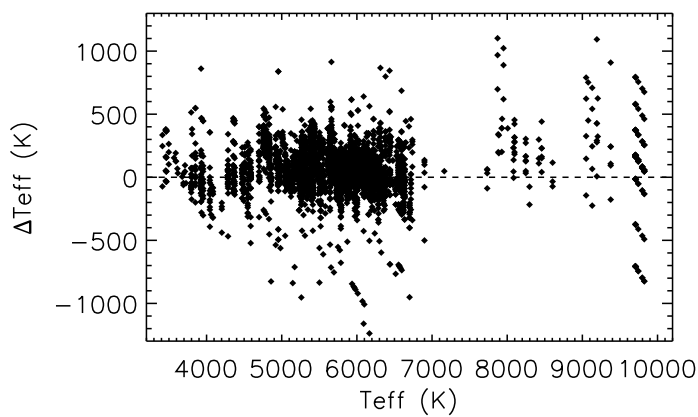}
\caption{Difference between fundamental effective temperatures  and other estimates listed in \pastel\ ($\Delta$\Teff = $T_{\rm other}-T_{\rm Teff}$) for 151 stars. The fundamental \Teff (x-axis) are based on direct measurements of angular diameters and total flux at  Earth. We note that several stars have up to six determinations of fundamental \Teff listed in \pastel.}
\label{f:fund_teff}
\end{figure}

 A new feature in the 2016 version of \pastel\ is to include several recent references providing effective temperatures from angular diameters and total flux at Earth for 151 stars \citep[261 determinations by ][]{boy12a, boy12b, boy13, von14,boy15, hei15}. They provide fundamental values of  \Teff, which define the absolute temperature scale. Figure \ref{f:fund_teff} displays the comparison of these fundamental values to the photometric or spectroscopic \Teff listed in \pastel. The fundamental \Teff are cooler by 51 K (median)  than the spectroscopic or photometric values, with a median absolute deviation (MAD) of 96 K. For most stars, the \Teff differences span at least 200-300 K. We have investigated the most extreme differences, i.e. those larger than 500 K. They mainly come from less reliable determinations in the earliest bibliographic references, or from the most uncertain determinations in \cite{gon09} who sometimes quote errors of several hundred K. Other large differences are difficult to explain without going into the details of the publications.

\subsection{Logarithm of surface gravity, \logg}
Surface gravities \logg are usually derived by  the method of ionization equilibrium or by the parallax method. A novelty of the 2016 version of \pastel\ is to have  \logg determinations based on asteroseismology. Several studies combining asteroseismology and spectroscopy are included  \citep[e.g.][]{cre12, tak15}.

The distribution of the 27\,178 \logg determinations is shown in Fig.~\ref{tgm}. A clear separation at \logg$\simeq$3.2 between the giants and the dwarfs and subgiants can be seen. Dwarfs and subgiants represent $\sim$65\% of the \logg determinations. 

When available in the papers, errors on \logg  are provided in \pastel. They have a median value of  0.10 dex, which agrees nicely with the median standard deviation (0.09 dex)  for stars having at least two \logg determinations available. 


\subsection{Metallicity, \FeH} 

Only \FeH determinations based on high-resolution, high signal-to-noise spectra are recorded in \pastel. High spectral resolution was considered to start at R=25\,000, and high S/N at 50. In a few cases, determinations that do not exactly fulfil the S/N requirements have been included. For instance, the high-resolution spectroscopic study of extremely metal-poor stars by \cite{aok13} has been entirely included even if a fraction of the spectra have a S/N lower than our limit owing to the very high interest of the targets. 

The distribution of the 26\,718 \FeH determinations is shown in Fig.~\ref{tgm}. The \FeH determinations range from -4.80 to +2.40. The solar metallicity is by far the most frequent value. More than 70\% of the \FeH determinations are between -0.50 and +0.50. 

Errors on \FeH  are also provided when available and have a median value of 0.06 dex. For the 5\,061 stars that have at least two \FeH  determinations, the median  standard  deviation is 0.05 dex. The median standard  deviation is even lower (0.04 dex) for FGK stars (4000 $\le$ \Teff $\le$ 6500 K) more metal-rich than \FeH=-1.0 dex. These good values of dispersion suggest that the errors quoted in the publications are in general not underestimated. For metal-poor FGK stars the median standard  deviation is 0.09 dex. A few stars (79) exhibit large discrepancies in \FeH determinations, with a standard deviation larger than 0.3 dex. These are mainly metal-poor stars and A-, B-, and M-type stars.

A flag in \pastel\ indicates when the metallicity determination relies on all metals (as often is the case for M dwarfs), or when [Fe/H] is corrected for  non-local thermodynamic equilibrium effects, or when [Fe/H] has been measured from FeII lines.

\begin{figure}[h!]
\begin{center}
\includegraphics[width=0.4\textwidth]{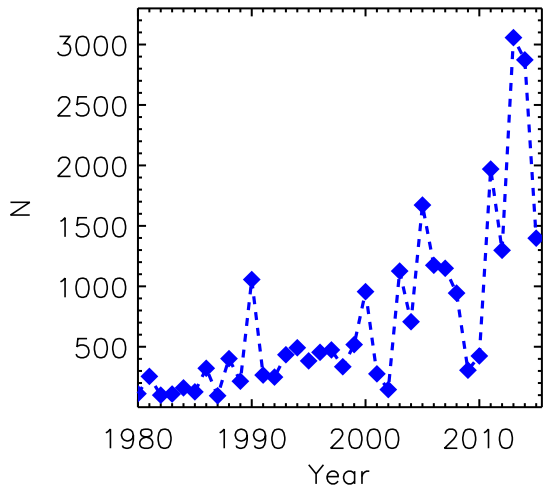}
\caption{Number of \FeH determinations per year of publication.}
\label{f:TF_yr}
\end{center}
\end{figure}

The methods of spectral analysis have evolved; they are more  automated and are applied to large numbers of spectra. Consequently, the number of \FeH determinations is rapidly increasing, as is shown in Fig. \ref{f:TF_yr}, which displays the number of \FeH determinations per year of publication. The increase is not regular, but significant. The number of studies involving hundreds of stars is also growing. Since 2010, eleven papers have published more than 250 \tgm determinations, and four papers present more than 500 stars \citep{ben14,luc14,ram13,sou11}. These four datasets do not overlap greatly, except \cite{ram13} and \cite{ben14}, which have 156 stars in common. These two samples agree very well in their metallicity scales, with a median metallicity difference of 0.02 dex (MAD=0.05 dex). 

\begin{figure*}[t]
\center
\includegraphics[width=0.9\textwidth]{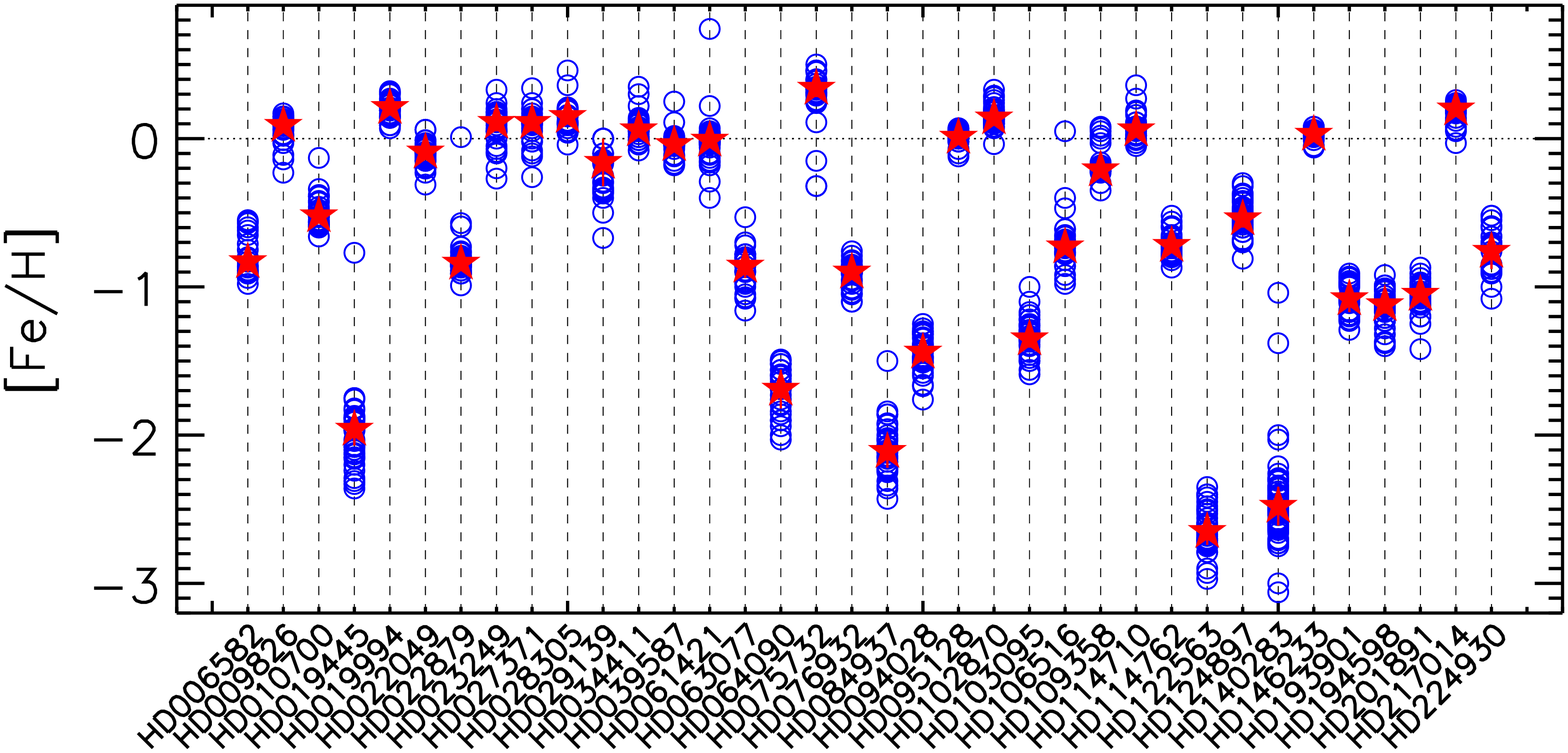}
\caption{Metallicities reported in \pastel\ for the 36 most studied stars listed in Table \ref{t:best_stars}  (open circles). The median value is represented by a red star.}
\label{f:comp_feh}
\end{figure*}

 Some stars in \pastel\ are very well studied, with \tgm determinations published by many different authors. As an illustration, we give in Table \ref{t:best_stars} the 36 stars that have at least 25 determinations available. The median parameters are provided together with the MAD and the number of determinations available for each parameter. The record is for the metal-poor subgiant HD140283 with nearly 60 determinations of metallicity, which present a substantial scatter (MAD=0.13 dex). The largest scatter (MAD=0.16 dex) is for the coolest star in this subsample, the giant HD029139 ($\alpha$ Tau). An excellent agreement between authors is obtained for several other stars, even if not of solar metallicity, like HD022879, a dwarf of metallicity \FeH=-0.84 (MAD=0.02 dex). Figure \ref{f:comp_feh} 
shows how metallicity varies from publication to publication. Visually, it is clear  that in most cases the \FeH determinations span about 0.5 dex owing to some discrepant determinations corresponding mostly to older publications. The users who need to average all the \FeH determinations available for one star should use the median value to avoid these outliers.

\begin{table*}
\centering
\caption{Median atmospheric parameters of the 36 most studied stars in PASTEL. The error is the median absolute deviation (MAD) obtained from the N determinations  available in the catalogue.}
\label{t:best_stars}
\begin{tabular}{lclclcl}
\hline 
HD     & \Teff &  N$_{\Teff}$ & \logg & N$_{\logg}$ &\FeH & N$_{\FeH}$   \\
\hline 
 HD006582&   5308$\pm$44 & 37  & 4.50$\pm$0.10 & 27 & -0.83$\pm$0.07 & 28 \\
 HD009826&   6155$\pm$51 & 39  & 4.14$\pm$0.08 & 28 & +0.09$\pm$0.04 & 29 \\
 HD010700&   5331$\pm$41 & 45  & 4.50$\pm$0.10 & 32 & -0.52$\pm$0.04 & 35 \\
 HD019445&   5985$\pm$74 & 48  & 4.39$\pm$0.11 & 40 & -1.96$\pm$0.11 & 43 \\
 HD019994&   6150$\pm$59 & 34  & 4.15$\pm$0.10 & 29 & +0.21$\pm$0.04 & 29 \\
 HD022049&   5084$\pm$51 & 37  & 4.54$\pm$0.08 & 30 & -0.09$\pm$0.05 & 29 \\
 HD022879&   5857$\pm$59 & 43  & 4.29$\pm$0.06 & 34 & -0.84$\pm$0.02 & 35 \\
 HD023249&   5023$\pm$77 & 30  & 3.80$\pm$0.04 & 25 & +0.11$\pm$0.06 & 28 \\
 HD027371&   4990$\pm$50 & 25  & 2.70$\pm$0.12 & 23 & +0.11$\pm$0.06 & 25 \\
 HD028305&   4915$\pm$69 & 27  & 2.70$\pm$0.13 & 24 & +0.15$\pm$0.05 & 25 \\
 HD029139&   3891$\pm$52 & 28  & 1.20$\pm$0.26 & 26 & -0.16$\pm$0.16 & 26 \\
 HD034411&   5860$\pm$30 & 35  & 4.22$\pm$0.04 & 22 & +0.06$\pm$0.05 & 25 \\
 HD039587&   5947$\pm$27 & 30  & 4.47$\pm$0.04 & 22 & -0.04$\pm$0.03 & 25 \\
 HD061421&   6582$\pm$68 & 53  & 4.00$\pm$0.02 & 36 & -0.01$\pm$0.03 & 43 \\
 HD063077&   5723$\pm$70 & 28  & 4.10$\pm$0.11 & 24 & -0.86$\pm$0.08 & 27 \\
 HD064090&   5400$\pm$38 & 34  & 4.50$\pm$0.20 & 27 & -1.69$\pm$0.13 & 27 \\
 HD075732&   5252$\pm$56 & 31  & 4.40$\pm$0.09 & 24 & +0.34$\pm$0.06 & 25 \\
 HD076932&   5860$\pm$40 & 37  & 4.10$\pm$0.10 & 30 & -0.90$\pm$0.04 & 32 \\
 HD084937&   6290$\pm$68 & 40  & 4.00$\pm$0.06 & 33 & -2.11$\pm$0.11 & 33 \\
 HD094028&   5995$\pm$61 & 38  & 4.19$\pm$0.13 & 29 & -1.44$\pm$0.07 & 30 \\
 HD095128&   5887$\pm$25 & 34  & 4.30$\pm$0.04 & 30 & +0.01$\pm$0.02 & 30 \\
 HD102870&   6103$\pm$43 & 43  & 4.14$\pm$0.06 & 29 & +0.14$\pm$0.04 & 31 \\
 HD103095&   5054$\pm$64 & 53  & 4.65$\pm$0.12 & 42 & -1.35$\pm$0.06 & 43 \\
 HD106516&   6208$\pm$61 & 36  & 4.40$\pm$0.10 & 26 & -0.73$\pm$0.05 & 28 \\
 HD109358&   5887$\pm$43 & 35  & 4.42$\pm$0.07 & 23 & -0.21$\pm$0.02 & 25 \\
 HD114710&   6000$\pm$50 & 38  & 4.43$\pm$0.06 & 22 & +0.06$\pm$0.04 & 28 \\
 HD114762&   5871$\pm$63 & 41  & 4.17$\pm$0.07 & 32 & -0.72$\pm$0.05 & 33 \\
 HD122563&   4583$\pm$17 & 46  & 1.20$\pm$0.20 & 39 & -2.65$\pm$0.09 & 42 \\
 HD124897&   4300$\pm$45 & 41  & 1.60$\pm$0.12 & 37 & -0.54$\pm$0.06 & 38 \\
 HD140283&   5690$\pm$65 & 63  & 3.58$\pm$0.13 & 54 & -2.48$\pm$0.13 & 57 \\
 HD146233&   5799$\pm$24 & 33  & 4.45$\pm$0.03 & 27 & +0.03$\pm$0.02 & 26 \\
 HD193901&   5745$\pm$82 & 31  & 4.46$\pm$0.12 & 26 & -1.08$\pm$0.09 & 26 \\
 HD194598&   5962$\pm$62 & 37  & 4.27$\pm$0.06 & 31 & -1.12$\pm$0.06 & 32 \\
 HD201891&   5900$\pm$47 & 38  & 4.31$\pm$0.09 & 32 & -1.05$\pm$0.05 & 32 \\
 HD217014&   5775$\pm$25 & 34  & 4.33$\pm$0.08 & 26 & +0.20$\pm$0.01 & 26 \\
 HD224930&   5362$\pm$87 & 26  & 4.40$\pm$0.18 & 20 & -0.76$\pm$0.12 & 25 \\
\hline
\end{tabular}
\end{table*}

There are 285 entries in \pastel\ that lack a \Teff determination because it  was not given in the corresponding paper. However, we still found it useful to add these metallicities in the catalogue, even without the corresponding \Teff. Some of these metallicities come from the original [Fe/H] catalogue by Cayrel de Strobel et al.

Abundances of  elements other than iron would be very useful. It is beyond the scope of \pastel\ to provide them, but we mention here several large compilations of abundances that can be cross-matched to \pastel: 
\begin{itemize}
\item the Hypatia Catalog \citep{hin14} made of 84 literature sources for 50 elements across 3058 stars in the solar neighbourhood within 150 pc of the Sun;
\item chemical abundances of 1111 FGK stars \citep{adi12};
\item homogeneous abundances from the literature for 743 stars \citep{sou05};
\item detailed elemental abundance study of 714 F and G dwarf stars in the solar neighbourhood \citep{ben14};
\item the SAGA database of extremely metal-poor stars \citep{sud08}.
\end{itemize}

\subsection{Bibliographical references}
The main astronomical journals and the CDS database have been surveyed to search for relevant publications. The 2016 version of \pastel\ includes papers until December 2015. For each publication introduced in \pastel, the name of the first author and the bibcode are given. 

Although \pastel\ is intended to be exhaustive, it is likely that we have missed some papers. We recommend users of the catalogue to notify us of any missing references that should be included in the next versions of the catalogue.

\section{Stellar content of the catalogue}

\begin{figure*}[h!]
\begin{center}
\includegraphics[width=0.8\textwidth]{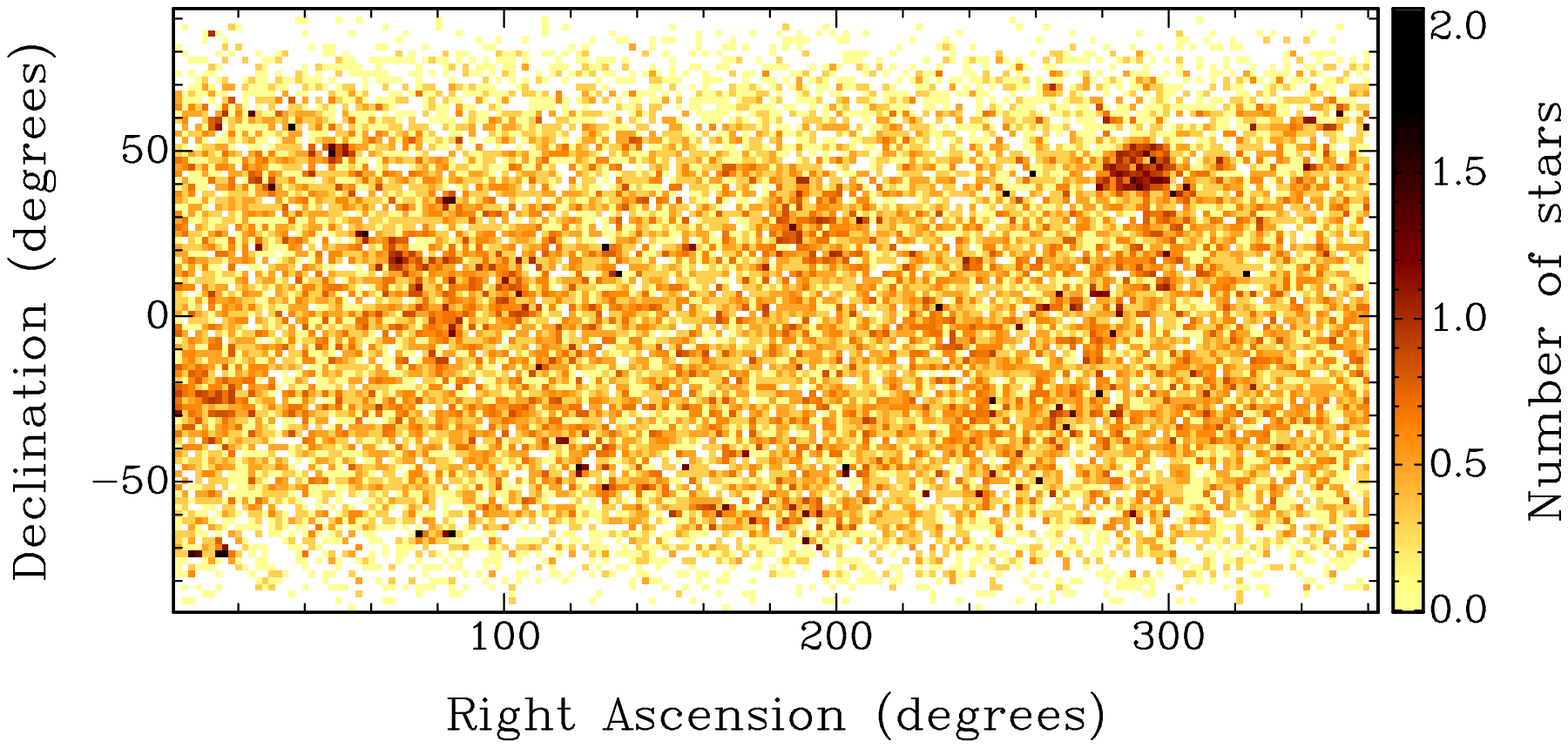}
\caption{Distribution of the  \pastel\ stars in equatorial coordinates. A logarithmic scale is used.}
\label{f:coord}
\end{center}
\end{figure*}

In Febuary 2016 \pastel\ included   31\,401 different stars (for 64\,082 records). 
 Fig.~\ref{f:coord} displays the distribution of  those stars in equatorial coordinates. The distribution is not perfectly homogeneous and the overdensities reflect the interest of astronomers in some peculiar fields, like the open clusters along the galactic plane, the Kepler field, or the galactic poles.
 
\begin{figure}[h!]
\begin{center}
\includegraphics[width=0.4\textwidth]{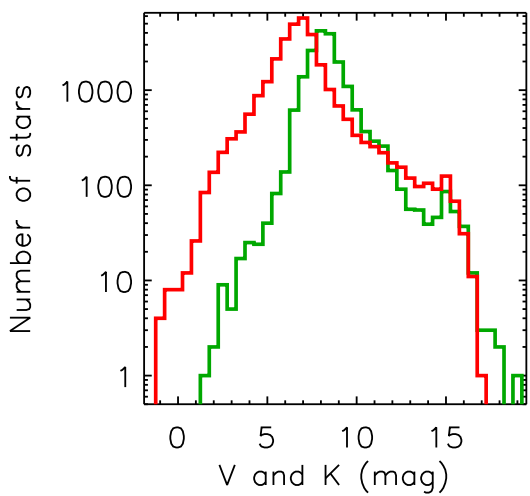}
\caption{Histograms of the V  magnitudes (green curve) and K magnitudes (red curve) of the \pastel\ stars.}
\label{f:VKmag}
\end{center}
\end{figure}

The histograms of V and K magnitudes are shown in Fig.~\ref{f:VKmag}. The V magnitude is not available for 546 stars, while the K magnitude is not available for 224 stars. \pastel\ is clearly dominated by bright stars, $\sim$86\% of them  brighter than V=10 and  $\sim$1\%  fainter than V=15. The fraction of faint stars has not increased since the last version of \pastel.

 There are  11\,197 stars in \pastel\ with the full set of atmospheric parameters. This number has almost doubled since  the 2010 version of the catalogue, which included 5\,954 such stars. 

\begin{figure*}[h!]
\center
\includegraphics[width=0.7\textwidth]{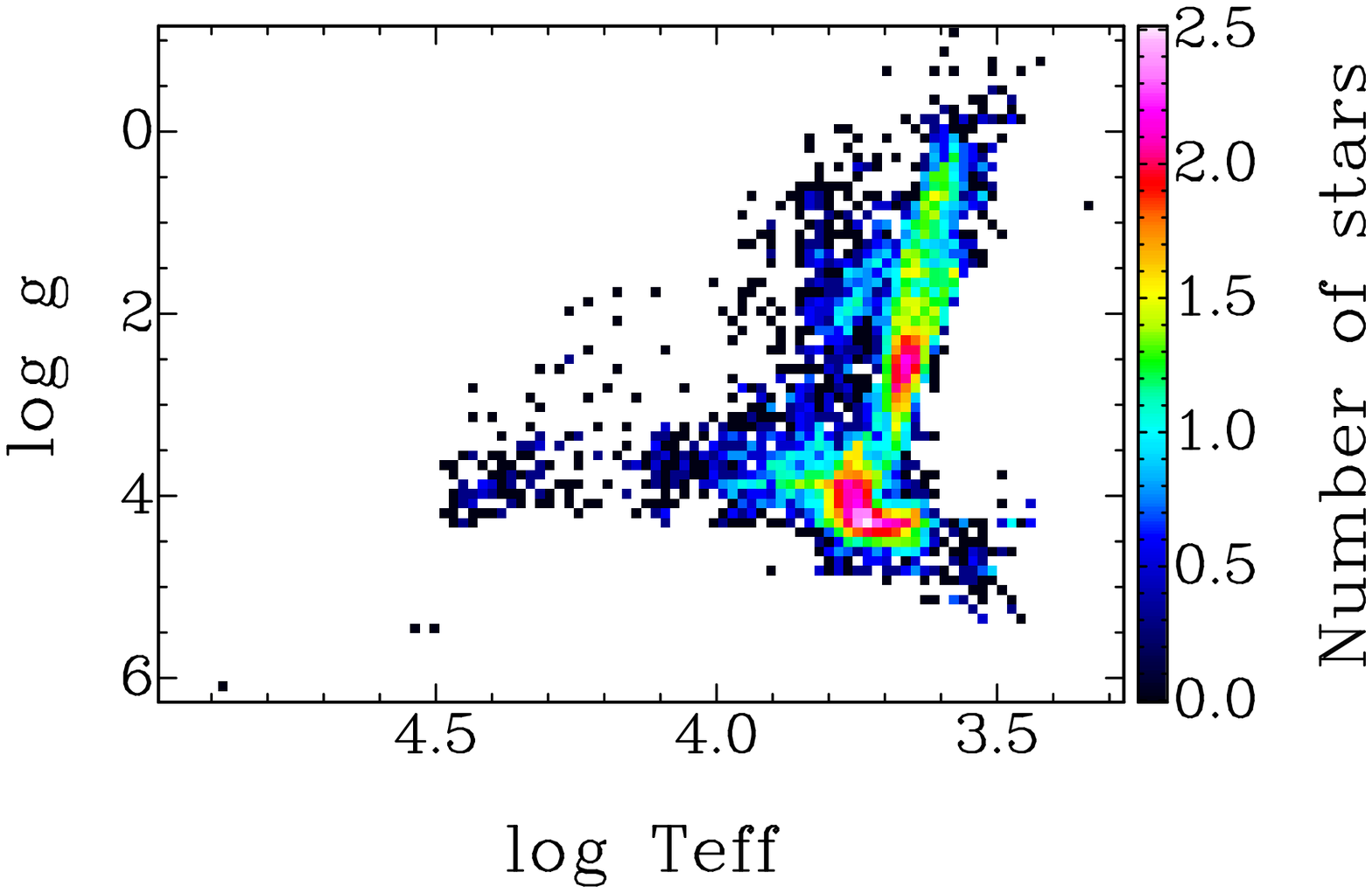}
\caption{Density plot of log(\Teff) / \logg  of the \pastel\ stars. A logarithmic scale is used.}
\label{logT_logg}
\end{figure*}

\begin{figure*}[h!]
\center
\includegraphics[width=0.4\textwidth]{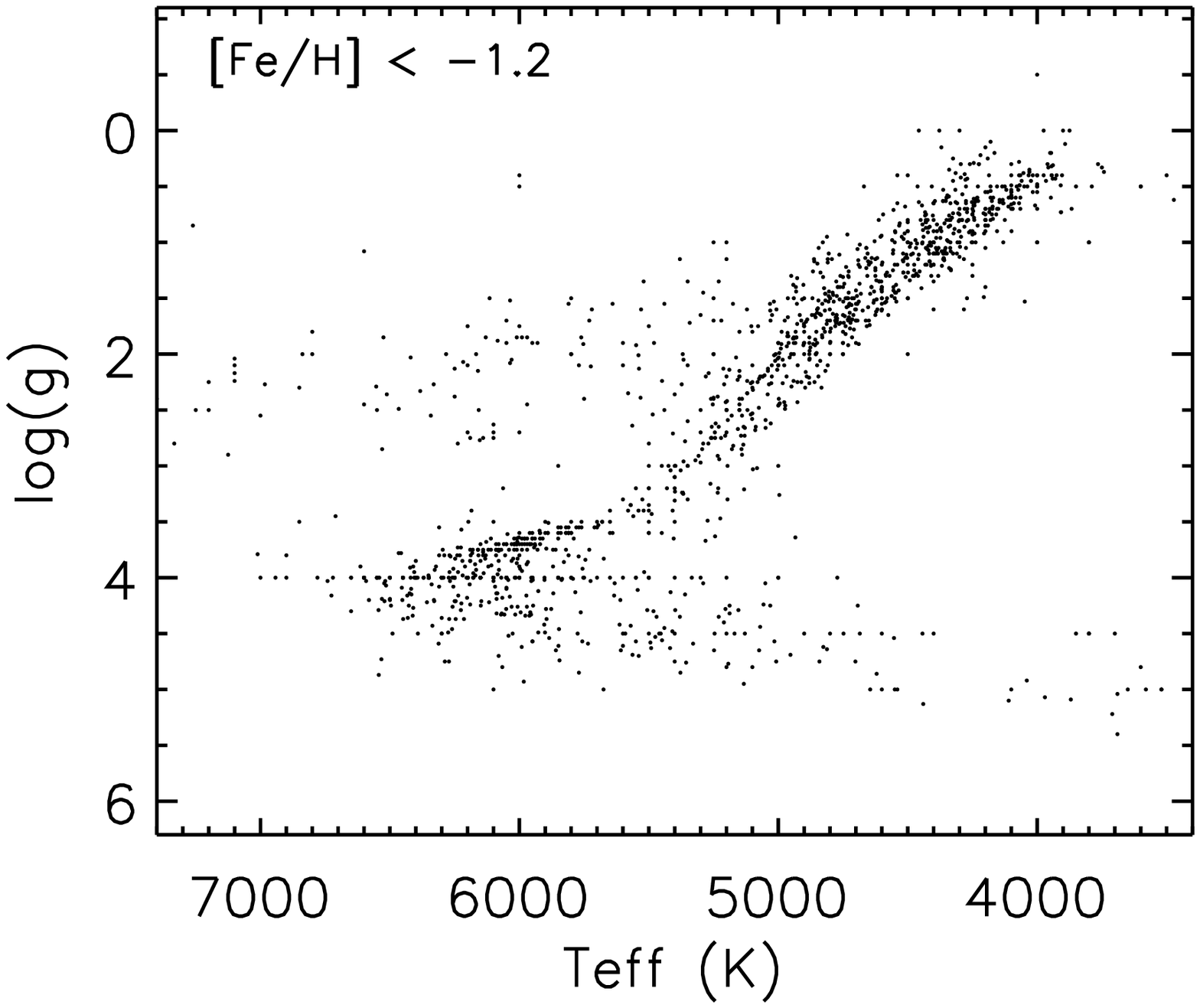}\includegraphics[width=0.4\textwidth]{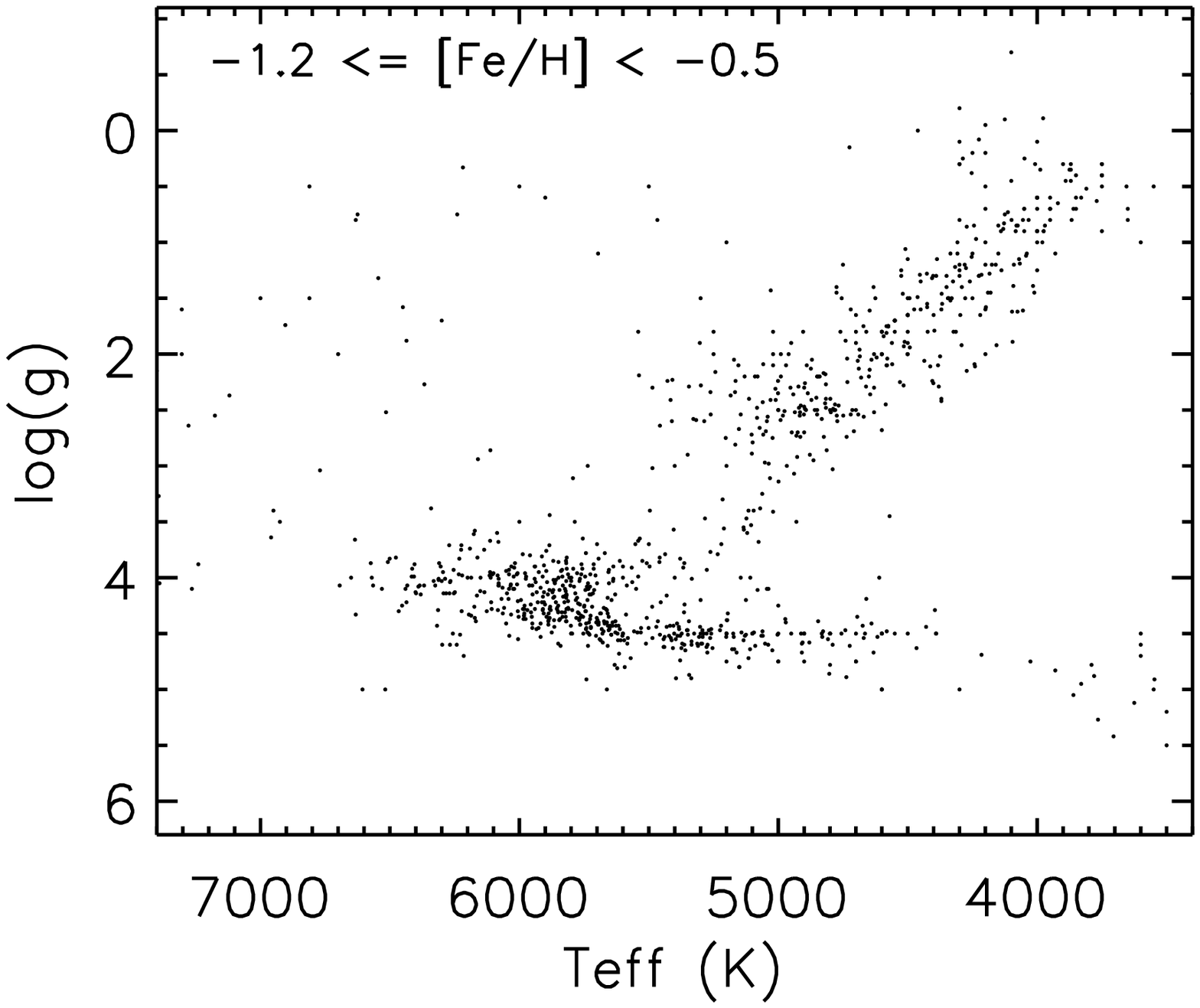}
\includegraphics[width=0.4\textwidth]{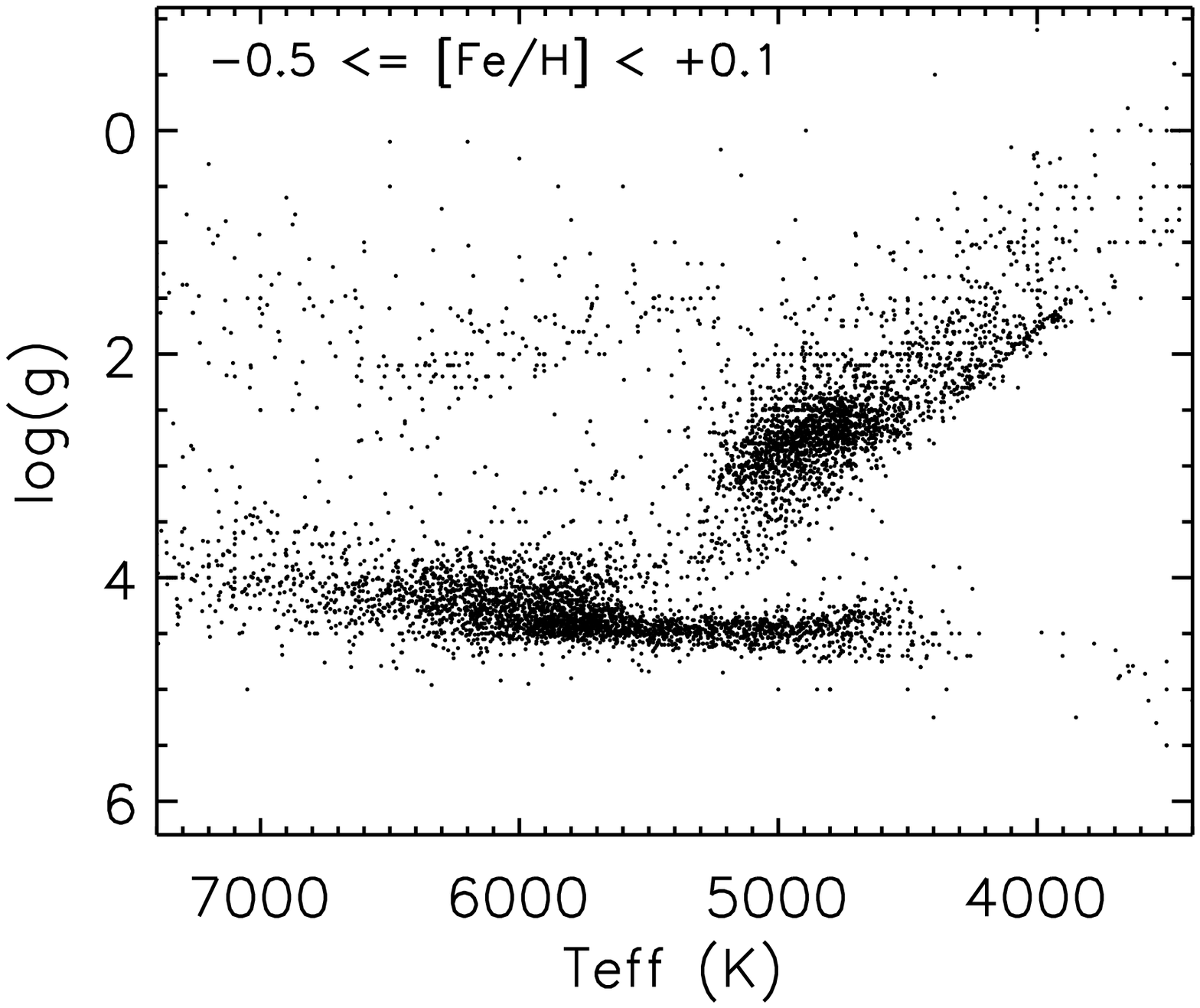}\includegraphics[width=0.4\textwidth]{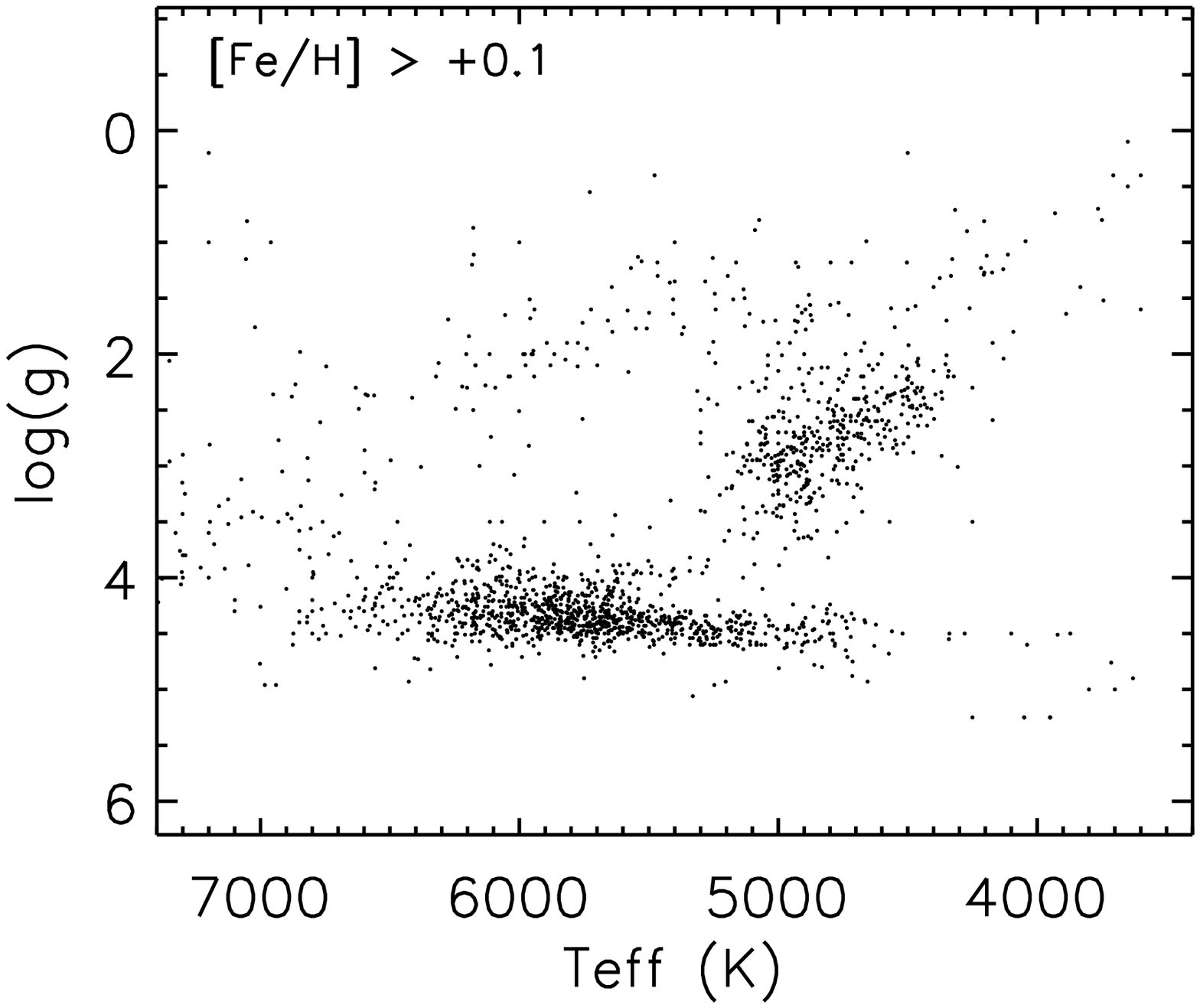}
\caption{\Teff / \logg diagram of the FGK stars in several regimes of metallicity corresponding roughly to the halo (top left), to the thick disc (top right), to the thin disc (bottom left), and to metal-rich stars (bottom right). For stars with several entries in \pastel,\ the median parameters have been adopted.}
\label{Teff_logg}
\end{figure*}

In order to show how these stars are distributed in \tgm the median parameters have been adopted for stars with several entries. Figure \ref{logT_logg} shows the density of the $\sim$11\,000 stars in the log\Teff / \logg plane, while Figure \ref{Teff_logg} shows the \Teff / \logg diagram in four bins of metallicity corresponding roughly to the halo, the thick disc, the thin disc, and super-metal rich stars. The stellar content of the catalogue  is not representative of the stellar content in the solar neighbourhood. It is biased towards stars that are massively studied in peculiar spectroscopic programmes, like solar-type stars in planet searches or metal-poor stars.

\section{Conclusion}

We have presented the new version of the \pastel\ catalogue, a bibliographical compilation of \tgm determinations. \pastel\ gathers stars that have been observed at  high spectral resolution and high S/N, and stars with a determination of \Teff based on various methods. The content of the catalogue has considerably increased since its 2010 original version. The number of entries has more than doubled  (30\,151 determinations in 2010 vs. 64\,082  today),  as has the number of stars with a determination of the three parameters \tgm (5\,954 stars in 2010 vs. 11\,197 stars today).

The methods used to measure atmospheric parameters are also evolving. A novelty of \pastel\ is to now include \Teff determined from the fundamental method, with angular diameters and total flux for 151 stars, as well as \logg determined from asteroseismology. The values of \tgm are also more often determined with automated methods applied to large samples involving hundreds of stars, which is  why the content of \pastel\ has  increased so much.

We have investigated how the errors of the parameters quoted in the literature compare with the dispersion observed from publication to publication. The agreement is good in general, showing that the errors in the literature are  essentially realistic. But there are also stars that exhibit  significant disagreement between spectroscopic analyses. This should lead the users of the catalogue to carefully analyse the references of the determinations they work with. The most reliable stars are those that have been studied by several different authors who agree on the parameters.  In general the determinations obtained before $\sim$1990 are less reliable because the spectra were registered on less efficient detectors, but we still found it useful to keep them in \pastel. It is recommended that  the median of the parameters be adopted instead of the mean if  a single value for a given star is needed. 

\pastel\ offers a  unique database for mining stars with known atmospheric parameters, and especially stars with a high-quality spectroscopic metallicity.   
 
\begin{acknowledgements}
We thank Alex Lobel who kindly notified us of a large number of references for hot stars.  We made extensive use of the CDS Simbad and VizieR services and of the NASA-ADS database, and we are  extremely grateful to the staff of these services for maintaining such valuable resources.
\end{acknowledgements}

\bibliography{28497}
\bibliographystyle{aa} 

\end{document}